\documentclass[conference]{IEEEtran}
\IEEEoverridecommandlockouts
\usepackage{cite}
\usepackage{amsmath,amssymb,amsfonts}
\usepackage{algorithmic}
\usepackage{graphicx}
\usepackage{textcomp}
\usepackage{booktabs}
\usepackage[ruled, vlined]{algorithm2e}
\usepackage{xcolor}
\DeclareMathOperator*{\argmax}{arg\,max} 
\renewcommand{\Re}{\operatorname{Re}}
\renewcommand{\Im}{\operatorname{Im}}
\def\BibTeX{{\rm B\kern-.05em{\sc i\kern-.025em b}\kern-.08em
    T\kern-.1667em\lower.7ex\hbox{E}\kern-.125emX}}

\usepackage{subcaption}

\usepackage{amsthm}
\newtheorem{theorem}{Theorem}
\newtheorem*{proof*}{Proof}

\usepackage{hyperref}
\makeatletter
\def\hyper@link#1#2#3{#3}%
\let\hyper@linkstart\@gobbletwo
\let\hyper@linkend\@empty
\makeatother

\begin{document}

\title{Robust Wideband Channel Estimation \\ for mmWave Massive MIMO Systems With \\ Beam Squint
\thanks{This work was supported in part by Huawei Research Project and in part by the National Science Foundation of China under Grants 61971279 and 62022054. (Corresponding author: Xue Jiang.)
}
}

\author{
Li Ge\textsuperscript{1}, Lin Chen\textsuperscript{1}, Xue Jiang\textsuperscript{1}, Weifeng Zhu\textsuperscript{2}, Qibo Qin\textsuperscript{3}, Xingzhao Liu\textsuperscript{1} \\ \space \\
\textsuperscript{1}School of Electronic Information and Electrical Engineering, Shanghai Jiao Tong University, Shanghai, China \\
\textsuperscript{2}Department of Electrical and Electronic Engineering, The Hong Kong Polytechnic University \\
\textsuperscript{3}Wireless Network RAN Research Department, Huawei Technologies Co. LTD, Shanghai, China\\
E-mail: \{geli2000an, linchenee, xuejiang, xzhliu\}@sjtu.edu.cn; eee-wf.zhu@polyu.edu.hk; qinqibo1@huawei.com
}

\maketitle

\begin{abstract}
This paper investigates the robust wideband channel estimation problem in the millimeter-wave (mmWave) massive multiple-input multiple-output (MIMO) systems. In such a scenario, the beam squint effect that the array response vectors vary with different frequencies and the impulsive noise are in existence, which pose great challenges for accurate channel estimation. Directly applying the existing channel estimation methods usually suffers significant performance degradation, since they are proposed based on the assumptions of frequency-invariant array response vectors and Gaussian distributed noise. To address these issues, this paper proposes a novel wideband channel estimation method with robustness to impulsive noise. Specifically, the proposed method incorporates a cyclic refinement step to overcome the estimation inaccuracy caused by the greedy nature of matching-pursuit-esque algorithms. In particular, the generalized $\ell_p$-norm minimization criterion is adopted in Newton's method to improve the performance robustness against the non-Gaussian impulsive noise. Numerical results are provided to verify the superior performance of the proposed method over the existing representative benchmarks.
\end{abstract}


\section{Introduction}
    \IEEEPARstart{W}{hen} it comes to developing future wireless communication networks, massive multiple-input multiple-output (MIMO) millimeter-wave (mmWave) communication has been considered one of the most promising technologies. Therein, the base station (BS) makes full use of a large number of antennas to enhance the spectral efficiency of multiple users under the same time-frequency resources. However, the signals received by different antennas actually have different time shifts\cite{swe_2}. Such asynchrony can become more noteworthy when the bandwidth becomes comparable to the carrier frequency, which is common in mmWave communication systems. When orthogonal frequency division multiplexing (OFDM) is being used for wideband transmissions, the beams observed by the receiver can deviate among different subcarriers. This phenomenon is known as beam squint effect\cite{BSE}.

    In wideband massive MIMO systems, the conventional MIMO channel model in \cite{conventional_channel_model_1,conventional_channel_model_2} usually cannot characterize the channels accurately due to the ignorance of the beam squint effect. Therefore, channel estimation methods designed in the past based on these models can suffer severe performance degradation in such wideband scenarios \cite{conventional_1,conventional_2}. Thus, early works usually concentrate on channel modeling of the wideband massive MIMO systems and then investigate their fundamentals. For example, \cite{bse_codebook} shows that the beam squint effect reduces the channel capacity with a specific case in which a uniform linear array (ULA) is adopted at the BS. In \cite{BoleiWang_grid_search}, the properties of the spatial wideband channel are comprehensively analyzed and a grid search-based channel estimation algorithm is proposed by exploiting the channel sparsities in the spatial and delay domain. Though properties of the wideband channel have been well disclosed, the proposed method suffers a fixed resolution constrained by the on-grid nature of the method. The authors in \cite{BoleiWang} and \cite{A_Block_Sparsity} propose the iterative reweighted approaches to handle the wideband channel estimation task. However, their methods use a large number of grids as initial candidate paths and decrease the number of paths in each iteration, which leads to high computational costs. On the other hand, the aforementioned works all concentrate on the wideband massive MIMO systems under the assumption that the background noise within the bandwidth satisfies the Gaussian distribution. In practical applications, such as smart grids\cite{power_impulsive_noise} and indoor wireless communications\cite{indoor_impulsive_noise}, the background noise is known to be impulsive or spiky. In the presence of such non-Gaussian noise, the widely used minimum mean squared error (MMSE) criterion-based channel estimation methods can experience significant performance deterioration \cite{Mitigation_techniques}. To the best of our knowledge, how to realize accurate channel estimation in the wideband mmWave massive MIMO systems with the beam squint effect and the non-Gaussian impulsive noise remains an open problem. This paper will address this issue.
    
    In this work, we consider the wideband channel estimation problem for the mmWave massive MIMO system with non-Gaussian noise and propose a robust wideband Newtonized orthogonal matching pursuit (wNOMP) channel estimation algorithm via $\ell_p$-norm minimization, which is a maximum likelihood (ML) estimation method under the generalized Gaussian distribution. By using the minimum $\ell_p$-norm criterion, the performance robustness against the impulsive noise can be improved. In the special case when $p=2$, the generalized $l_p$-norm criterion reduces to the commonly used MMSE criterion, which is optimal for the scenarios with Gaussian noise. On the other hand, the proposed wNOMP algorithm contains a cyclic refinement step using Newton's method to overcome the estimation inaccuracy caused by the greedy nature of MP-type algorithms with negligible computational complexity increase. We have also established the convergence results of the proposed method. Experiment results are provided to verify the superiority of the proposed algorithm.
    
    The rest of this paper is organized as follows. Section \uppercase\expandafter{\romannumeral2} introduces the system model. Section \uppercase\expandafter{\romannumeral3} presents the proposed wideband channel estimation algorithm induced by generalized Gaussian noise. Convergence results are established in Section \uppercase\expandafter{\romannumeral4}. Experiment results and complexity are contained in Section \uppercase\expandafter{\romannumeral5}. Section \uppercase\expandafter{\romannumeral6} concludes the paper.
    
    \textit{Notations}: Uppercase and lowercase boldface denote matrices and vectors, respectively; transpose, conjugate, conjugate transpose, inverse, and Moore-Penrose pseudoinverse of matrix $\mathbf{A}$ are written as $\mathbf{A}^\text{T}, \mathbf{A}^\text{*}, \mathbf{A}^\text{H}, \mathbf{A}^{-1}, \mathbf{A}^{\dag}$, respectively; $\Vert \cdot \Vert_p$ and $\Vert \cdot \Vert_{\text{F}}$ denote the $\ell_p$-norm and the Frobenius norm, respectively; $\mathbf{A}_{i, j}$ stands for the $(i, j)$th element of $\mathbf{A}$; $\text{diag}\{\mathbf{a}\}$ denotes a diagonal matrix with the entries in $\mathbf{a}$ filling its diagonal line and $\text{diag}\{\{\mathbf{A}_m\}_{m=1}^{M}\}$ stands for a block diagonal matrix with $\{\mathbf{A}_m\}_{m=1}^{M}$ being its diagonal submatrices in order; $\Re\{\cdot\}$ and $\Im\{\cdot\}$ denote the real and imaginary part of a complex number, vector or matrix; $\odot$ and $\otimes$ are the Hadamard product and the Kronecker product of matrices, respectively; $\bar{a}$ denotes the normalized equivalence of scalar variable $a$ and $\hat{a}$ denotes its estimation. 

\section{System Model} 
        \begin{figure*}[t]
            \centering
            \begin{subfigure}{0.48\textwidth}
                \includegraphics[width=\textwidth, height=0.43\textwidth]{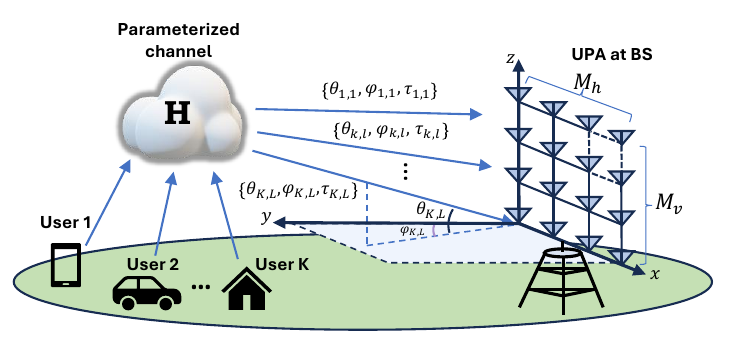}
                \caption{Uplink transmission of $K$ users.}
                \label{fig:uplink_transmission}
            \end{subfigure}
            \begin{subfigure}{0.48\textwidth}
                \includegraphics[width=\textwidth, height=0.39\textwidth]{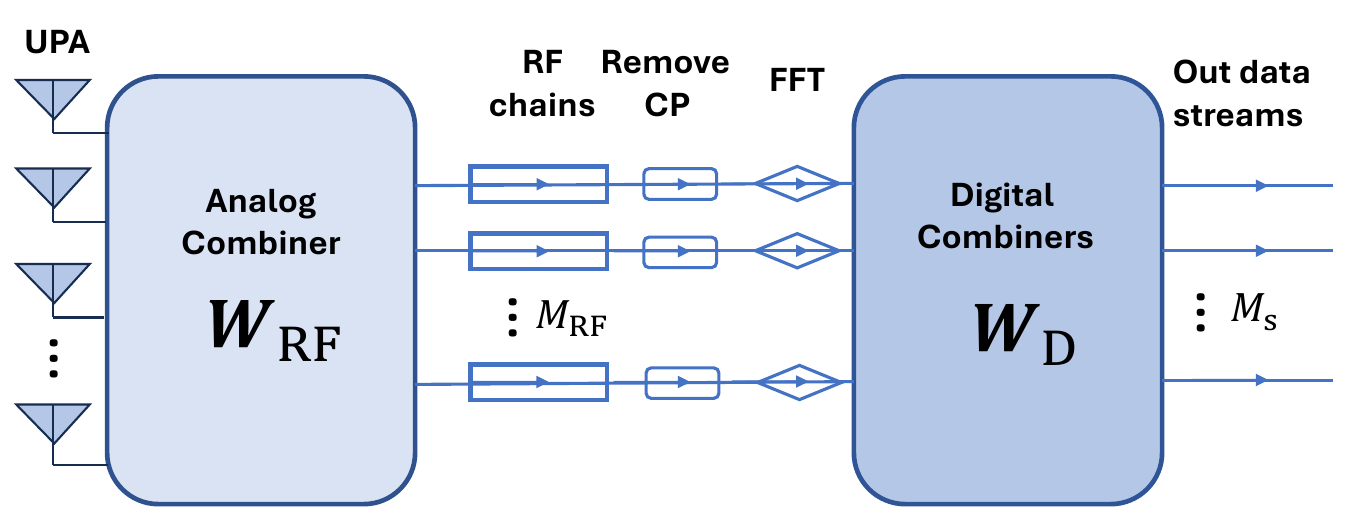}
                \caption{Hybrid processing at the BS.}
                \label{fig:hybrid_architecture}
            \end{subfigure}
            \caption{An uplink transmission scenario with a hybrid architecture at the BS.}
        \end{figure*}
        We consider an uplink transmission scenario as illustrated in Fig. \ref{fig:uplink_transmission}, assuming that the BS is equipped with a uniform planar array (UPA), containing $M_h$ antennas arranged horizontally and $M_v$ antennas arranged vertically. The BS serves $K$ users randomly distributed in a cell. Each user is equipped with a single antenna.
        
        Assuming that the received signal propagates through $L$ paths, $\theta_{k, l} \in (0, \pi)$ and $\varphi_{k, l} \in (-\pi, \pi)$ are defined to be the elevation angle and the azimuth of the $l$th path of the $k$th user. According to the geometric model, the delay of the $l$th path between the antenna of the $k$th user and the $(v, h)$th (vertical and horizontal) antenna of the BS can be expressed as
        \begin{small}
        \begin{align}
            \tau_{k, l, v, h} = \tau_{k, l} + \frac{(v-1)d_v\sin\theta_{k, l}}{c} + \frac{(h-1)d_h\cos\theta_{k, l}\sin\varphi_{k, l}}{c},
            \label{eq:delay}
        \end{align}
        \end{small}
        where $\tau_{k, l}$ is defined as the delay of the $l$th path between the antenna of the $k$th user and the $(1, 1)$th antenna of the BS, $d_v$ and $d_h$ stand for the vertical and horizontal antenna spacing, respectively, and $c$ is the speed of light.
        
        Defining $f_c$ and $\lambda_c$ to be the frequency and the wavelength of the carrier, we can express the channel in the frequency domain as
        \begin{align}
            h_{k, l, v, h}(f) = g_{k, l} ~ e^{-j2\pi f \tau_{k, l, v, h}} e^{-j2\pi f_c \tau_{k, l, v, h}},
            \label{eq:channel_1}
        \end{align}
        where $g_{k, l}$ stands for the gain of the $l$th path of the $k$th user. 
        
        Assuming $d_v \negthickspace = d_h = \lambda_c/2$, plug \eqref{eq:delay} into \eqref{eq:channel_1} and define $\bar{g}_{k, l} = g_{k, l}~e^{-j2\pi f_c \tau_{k, l}}$ as the equivalent complex propagation gain, we can stack $h_{k, l, v, h}(f)$ in the two dimension of antennas to obtain a matrix form of the channel as
        \begin{align}
            \mathbf{H}_{k, l}(f) = \bar{g}_{k, l} e^{-j2\pi f \tau_{k, l}}\mathbf{a_v}(\bar{\theta}_{k, l}, f) {\mathbf{a_h}}^\text{T}(\bar{\varphi}_{k, l}, f) \in \mathbb{C}^{M_v \times M_h},
            \label{eq:channel_matrix}
        \end{align}
        where
        \begin{gather}
            \mathbf{a_v}(\bar{\theta}_{k, l}, f) = [1, e^{-j\pi(1+\frac{f}{f_c})\bar{\theta}_{k, l}}, \hdots, e^{-j\pi(M_v-1)(1+\frac{f}{f_c})\bar{\theta}_{k, l}}]^{\text{T}},\\
            \mathbf{a_h}(\bar{\varphi}_{k, l}, f) = [1, e^{-j\pi(1+\frac{f}{f_c})\bar{\varphi}_{k, l}}, \hdots, e^{-j\pi(M_h-1)(1+\frac{f}{f_c})\bar{\varphi}_{k, l}}]^{\text{T}},
        \end{gather}
        and $\bar{\theta}_{k, l} = \sin{\theta_{k, l}}, \bar{\varphi}_{k, l} = \cos{\theta_{k, l}}\sin{\varphi_{k, l}}$ are the normalized elevation angle and azimuth.
        
        Note that the steering vectors $\mathbf{a_v}(\bar{\theta}_{k, l}, f)$ and $\mathbf{a_h}(\bar{\varphi}_{k, l}, f)$ are frequency dependent, which stands for beam squint effect\cite{BSE}. When the number of antennas is relatively small or a narrowband scenario is considered, the beam squint term $e^{-j\pi\frac{m_v f}{f_c}\bar{\theta}_{k, l}}$ and $e^{-j\pi\frac{m_h f}{f_c}\bar{\varphi}_{k, l}}$ can be ignored. However, when considering a wideband MIMO scenario, the baseband frequency $f$ can be comparable to $f_c$ and the number of antennas can be large, the beam squint effect cannot be neglected.
        
        We assume OFDM is being used with $N$ subcarriers over the bandwidth $B$ for all of the $K$ users. Let each user be assigned with $T=N/K$ subcarriers and the subcarrier spacing is $\Delta f$. Donate $\mathcal{N}_k$ to be the index set of the subcarriers occupied by the $k$th user, such that $\bigcup_{k=1}^{K} \mathcal{N}_k = \left\{0, 1, \hdots, N-1\right\}, \mathcal{N}_i \cap \mathcal{N}_j = \emptyset, i \neq j$. 
                
        A quantized phase-shifter-based hybrid architecture like \cite{tcas2} is considered for its practicality in mmWave communication systems as shown in Fig. \ref{fig:hybrid_architecture}. Donate $\mathbf{W}_{{\text{D}}, k, t} \in \mathbb{C}^{M_s \times M_{\text{RF}}}, \mathbf{W}_{{\text{RF}}, k} \in \mathbb{C}^{M_{\text{RF}} \times M_vM_h}$ to be the digital and analog combiners of the $t$th subcarrier of the $k$th user, where $M_s, M_{\text{RF}}$ are the number of data streams and RF chains, respectively. The phases of the entries in $\mathbf{W}_{{\text{RF}}, k}$ are randomly drawn from $\mathcal{A}_Q = \left\{ 2\pi \frac{i}{2^Q}, i =  0, 1, \hdots \hdots, 2^Q - 1 \right\}$, where $Q$ is the number of the quantization bits.
        
        We stack the channel in the dimension of subcarriers and perform vectorization. A single channel codeword  $\mathbf{c} \in \mathbb{C}^{M_v M_h T \times 1}$ can be written as
        \begin{equation}
        \begin{aligned}
            \mathbf{c}(\bar{\theta}, \bar{\varphi}, \tau, \mathcal{N}) = &    [\mathbf{a_h}^\text{T}(\bar{\varphi}, n_1\Delta f) \otimes \mathbf{a_v}^{\text{T}}(\bar{\theta}, n_1\Delta f) e^{-j2\pi n_1\Delta f \tau}, \\
             \hdots, \mathbf{a_h}^\text{T}&(\bar{\varphi}, n_T\Delta f) \otimes \mathbf{a_v}^{\text{T}}(\bar{\theta}, n_T\Delta f) e^{-j2\pi n_T\Delta f \tau}]^\text{T},
        \end{aligned}
        \label{eq:codeword}
        \end{equation}
        where $n_t \in \mathcal{N}, t = \left\{1, 2, \hdots, T\right\}$. It can be clearly observed from  \eqref{eq:codeword} that beam squint effect extinguishes the low-rank property of the channel in the subcarrier dimension, thus causing a performance degradation of methods that utilize multiple subcarriers to perform joint estimation.
        
        Since the sets of the subcarriers occupied by each single antenna user are non-intersect, we can assume all users send "1" over their subcarriers for simplicity. By \eqref{eq:channel_matrix} and \eqref{eq:codeword}, the received signal can be expressed as
        \begin{equation}
            \mathbf{y}_k = \mathbf{W}_k \sum_{l=1}^{L}\bar{g}_{k, l}~\mathbf{c}(\bar{\theta}_{k, l}, \bar{\varphi}_{k, l}, \tau_{k, l}, \mathcal{N}_k) + \mathbf{z}_k = \mathbf{W}_k \mathbf{h}_k + \mathbf{z}_k,
            \label{eq:channel}
        \end{equation}
        where $\mathbf{W}_{k} = \text{diag}\left\{ \mathbf{W}_{{\text{D}}, k, 1}\mathbf{W}_{{\text{RF}}, k}, \hdots,  \mathbf{W}_{{\text{D}}, k, T}\mathbf{W}_{{\text{RF}}, k} \right\} \in \mathbb{C}^{M_sT \times M_vM_hT}$ is a block diagonal matrix and $\mathbf{z}_k$ is the additive complex noise. Our goal is to recover the uplink channel vector $\mathbf{h}_k$ from the noisy received signal $\mathbf{y}_k$.
        
\section{Robust Wideband Channel Estimation}
    To ensure the accuracy of channel estimation in real communication systems, we consider the complex generalized Gaussian noise (CGGN) model, which can well carve the characteristics of both common Gaussian noise and impulsive noise. To tackle the problem introduced by the aforementioned beam squint effect, we propose a robust wNOMP channel estimation algorithm inspired by the NOMP algorithm \cite{NOMP}. As this section will shown, the CGGN model induces a minimum $\ell_p$-norm criterion, which facilitates the robustness of the proposed wNOMP algorithm.

    \subsection{CGGN Model}
            We briefly introduce the generalized Gaussian distribution for the development of the proposed algorithm. Consider a complex generalized Gaussian random variable $Z$ with $\Re\{Z\}$ and $\Im\{Z\}$ being linearly uncorrelated, the probability density function (PDF) of $Z$ can be expressed as
            \begin{equation}
            \begin{aligned}
                f(z, \sigma, p) = & \frac{\beta(\frac{p}{2})}{\sigma^2} e^{-\frac{(2\eta(\frac{p}{2}))^{\frac{p}{2}}}{\sigma^p}\vert z \vert^p},
            \end{aligned}
            \label{eq:CGGN}
            \end{equation}
            where $\beta(p) = \frac{p\Gamma(\frac{4}{p})}{\pi\Gamma(\frac{2}{p})^2},~\eta(p) = \frac{\Gamma(\frac{4}{p})}{2\Gamma(\frac{2}{p})}$, $\Gamma(\cdot)$ is the Gamma function, $\sigma$ stands for the standard deviation and $p$ determines how heavy the tail of the PDF is. When $p = 2$, it is exactly the Gaussian distribution. When $p<2$ or $p>2$, the CGGN becomes more or less heavy-tailed than that of the Gaussian distribution. More details of CGGN can be found in \cite{CGG}.

    \subsection{Robust Wideband Channel Estimation}
        We consider the model in \eqref{eq:channel} with $z_k$ being CGGN as \eqref{eq:CGGN}, the log-likelihood function can be expressed as
        \begin{small}
        \begin{equation}
            \begin{aligned}
                \mathcal{L} & = \log\left(\prod_{i=1}^{M_vM_hT} \frac{\beta(\frac{p}{2})}{\sigma^2} e^{-\frac{(2\eta(\frac{p}{2})^{\frac{p}{2}}\vert {(\mathbf{y}_k - \mathbf{W}_k \sum_{l=1}^{L_c} {\hat{g}}_{k, l} \mathbf{c}_{k, l})}_i {\vert}^p}{\sigma^p}} \right)\\
                & = \sum_{i=1}^{M_vM_hT} \log\frac{\beta(\frac{p}{2})}{\sigma^2} - \frac{(2\eta(\frac{p}{2})^{\frac{p}{2}}}{\sigma^p} \vert (\mathbf{y}_k - \mathbf{W}_k \sum_{l=1}^{L_c} {\hat{g}}_{k, l} \mathbf{c}_{k, l})_i {\vert}^p,
            \end{aligned}
            \label{eq:log_likelihood}
        \end{equation}
        \end{small}
        where $L_c$ is the number of currently detected paths. After removing constants, maximization of the log-likelihood is equivalent to minimizing the $\ell_p$-norm of the residual:
        \begin{align}
         \min_{\hat{\theta}_{k, l}, \hat{\phi}_{k, l}, \hat{\tau}_{k, l}, \hat{g}_{k, l}} ~ \Vert \mathbf{y}_k - \mathbf{W}_k \sum_{l=1}^{L_c} \hat{g}_{k, l} \mathbf{c}(\hat{\theta}_{k, l}, \hat{\varphi}_{k, l}, {\hat{\tau}}_{k, l}, \mathcal{N}_k) \Vert_p^p.
         \label{eq:minimize_p-norm}
        \end{align}
        When $p=2$, CGGN regresses to the complex Gaussian noise (CGN) and \eqref{eq:minimize_p-norm} regresses to the commonly used MMSE criterion.

        To solve \eqref{eq:minimize_p-norm}, we need to determine the parameter set $\mathcal{P}_k = \left\{(\hat{\theta}_{k, l}, \hat{\varphi}_{k, l}, \hat{\tau}_{k, l})\right\}_{l=1}^{L_c}$ and corresponding path gain $\mathbf{\hat{g}}_{k}$. One can note that, once $\mathcal{P}_k$ is given, the optimal $\mathbf{\hat{g}}_{k}$ can be obtained by iterative weighted least square (IWLS):
        \begin{align}
            \mathbf{\hat{g}}_{k} = {((\mathbf{W}_k\mathbf{X})^{\text{H}}\mathbf{M}\mathbf{W}_k\mathbf{X})}^{-1} (\mathbf{W}_k\mathbf{X})^{\textbf{H}}\mathbf{M}\mathbf{y}_k,
            \label{eq:WLS}
        \end{align}
        where
        \begin{align}
            \mathbf{M} = \text{diag}\left\{ {\vert \mathbf{y}_k - \mathbf{W}_k\mathbf{X} \mathbf{\hat{g}}_k \vert}^{p-2} \right\}
            \label{eq:M}
        \end{align}
        is a diagonal weighting matrix and $\mathbf{X} = [\mathbf{c}_{k, 1}, \hdots, \mathbf{c}_{k, L_c}]$. Note that in order to determine $\mathbf{M}$, $\mathbf{\hat{g}}_k$ itself is required, which can be realized by an iterative process. To be specific, we use the least square estimation ${((\mathbf{W}_k\mathbf{X})^{\text{H}}\mathbf{W}_k\mathbf{X})}^{-1} (\mathbf{W}_k\mathbf{X})^{\textbf{H}}\mathbf{y}_k$ as the initial value of $\mathbf{\hat{g}}_{k}$ then iterate between \eqref{eq:WLS} and  \eqref{eq:M} until the interval of $\mathbf{\hat{g}}_{k}$ is sufficiently small. The elaborate derivation of \eqref{eq:WLS} and \eqref{eq:M} can be found in Appendix.

        The parameter set $\mathcal{P}_k$ is filled stepwise and refined in iterations, which are achieved by two operations: the greedy OMP step, and the refinement. In the $l$th iteration, we add the tuple $(\hat{\theta}_{k, l}, \hat{\varphi}_{k, l}, \hat{\tau}_{k, l})$ by minimizing
        \begin{align}
            S(\mathbf{y}_k^{(l-1)}, g, \mathbf{c}_{k}) = 
            \Vert {\mathbf{y}_k}^{(l-1)} - \mathbf{W}_k g \mathbf{c}(\theta, \varphi, \tau, \mathcal{N}_k) \Vert_p^p,
            \label{eq:min_l_p_each_iteration}
        \end{align}
        where $\mathbf{y}_k^{(l-1)}$ is the residual of the last iteration and $\mathbf{y}_k^{(0)} =\mathbf{y}_k$. Next, we explain these two steps in detail.

        \noindent\textbf{OMP step:} First, we choose a candidate path $\mathbf{c}_{k, l}$ from a pre-generated codebook
         \begin{align*}
            &\mathcal{C}(\hat{\mathbf{\theta}}, \hat{\mathbf{\varphi}}, \hat{\mathbf{\tau}}, \mathcal{N}_k)=\\
            &\left\{\mathbf{c}(\hat{\theta}, \hat{\varphi}, \hat{\tau}, \mathcal{N}_k) \left|
            \begin{scriptsize}
                \begin{aligned}
                    \hat{\theta} \in & \left\{0, \frac{1}{N_{\hat{\theta}}}, \hdots, \frac{N_{\hat{\theta}} - 1}{N_{\hat{\theta}}}\right\} \\
                    \hat{\varphi} \in & \left\{-1,\frac{2-N_{\hat{\varphi}}}{N_{\hat{\varphi}}}, \hdots, \frac{N_{\hat{\varphi}}-2}{N_{\hat{\varphi}}}\right\} \\
                    \hat{\tau} \in & \left\{0, \frac{\tau_m}{N_{\hat{\tau}}}, \hdots, (1 - \frac{1}{N_{\hat{\tau}}})\tau_m\right\}
                \end{aligned}
            \end{scriptsize}
            \right.
            \right\},
         \end{align*}
        where $N_{\hat{\theta}}, N_{\hat{\varphi}}, N_{\hat{\tau}}$ are the dimensions of the codebook and $\tau_m$ is the maximum propagation delay. Since calculating path gain for all candidate paths in $\mathcal{C}(\hat{\mathbf{\theta}}, \hat{\mathbf{\varphi}}, \mathbf{\tau}, \mathcal{N}_k)$ by  \eqref{eq:WLS} and \eqref{eq:M} to attain the residual can be computationally prohibitive, we choose the most correlative codeword to reduce the complexity
        \begin{align}
            \mathbf{c}({\hat{\theta}}_{k, l}, {\hat{\varphi}}_{k, l}, \hat{\tau}_{k, l}, \mathcal{N}_k) = & \argmax_{\mathbf{c}(\hat{\theta}, \hat{\varphi}, \hat{\tau}, \mathcal{N}_k)} \frac{\vert \mathbf{c}^{\text{H}}(\hat{\theta}, \hat{\varphi}, \hat{\tau}, \mathcal{N}_k) {\mathbf{W}_k}^{\text{H}} \mathbf{y}_k^{(l)}\vert}{{\Vert \mathbf{W}_k\mathbf{c}(\hat{\theta}, \hat{\varphi}, \hat{\tau}, \mathcal{N}_k)\Vert}_2^2},
            \label{eq:matching}
        \end{align}
        and obtain $\hat{g}_{k, l}$ according to \eqref{eq:WLS} and \eqref{eq:M}.

    \noindent\textbf{Refinement:} After the on-grid coarse detection, we use Newton's method to perform refinement as
    \begin{align}
        &
        \begin{pmatrix}
            \hat{\theta}_{k, l}^{\text{re}}, \hat{\varphi}_{k, l}^{\text{re}}, \hat{\tau}_{k, l}^{\text{re}}
        \end{pmatrix}^{\text{T}} = 
        \begin{pmatrix}
            \hat{\theta}_{k, l}, \hat{\varphi}_{k, l}, \hat{\tau}_{k, l}
        \end{pmatrix}^{\text{T}} - \nonumber\\& 
        \begin{pmatrix}
            \frac{\partial^2 S}{\partial {\hat{\theta}_{k, l}}^2} & \frac{\partial^2 S}{\partial \hat{\theta}_{k, l} \partial \hat{\varphi}_{k, l} } & \frac{\partial^2 S}{\partial \hat{\theta}_{k, l} \partial \hat{\tau}_{k, l}}\\ \frac{\partial^2 S}{\partial \hat{\theta}_{k, l} \partial \hat{\varphi}_{k, l} } &  \frac{\partial^2 S}{\partial {\hat{\varphi}_{k, l}}^2} & \frac{\partial^2 S}{\partial \hat{\varphi}_{k, l} \partial \hat{\tau}_{k, l}}\\ \frac{\partial^2 S}{\partial \hat{\theta}_{k, l} \partial \hat{\tau}_{k, l}} & \frac{\partial^2 S}{\partial \hat{\varphi}_{k, l} \partial \hat{\tau}_{k, l}} & \frac{\partial^2 S}{\partial {\hat{\tau}_{k, l}}^2}
        \end{pmatrix}^{-1}
        \begin{pmatrix}
            \frac{\partial S}{\partial \hat{\theta}_{k, l}} \\ \frac{\partial S}{\partial \hat{\varphi}_{k, l}} \\ \frac{\partial S}{\partial \hat{\tau}_{k, l}}
        \end{pmatrix}_.
        \label{eq:Newton's_method}
    \end{align}
    
    Here, we show the results of the derivatives and the details of their derivation are elaborated in Appendix.
    \begin{align}
        \centering
        \label{eq:derivative_1} \frac{\partial S}{\partial \hat{\theta}_{k, l}} & = 2\Re{\left\{ \frac{p\cdot\hat{g}_{k, l}}{2} {{\mathbf{y}_{k}}^{(l)}}^{\text{H}} \mathbf{M} \mathbf{W}_k \frac{\partial \mathbf{c}_{k, l}}{\partial \hat{\theta}_{k, l}}\right\}}_, 
    \end{align}
    \vspace{-0.3cm}
    \begin{align}
        \label{eq:derivative_2}\frac{\partial^2 S}{\partial \hat{\theta}_{k, l}^2} & = 
        2\Re \left\{\frac{p(p-2)}{4} {\hat{g}_{k, l}}^2 \left[ ({{\mathbf{y}_{k}}^{(l)}}^{\text{H}})^2 \odot \vert {{\mathbf{y}_{k}}^{(l)}}^{\text{H}} \vert^{p-4} \right] \right. \notag\\
           & {(\mathbf{W}_k \frac{\partial \mathbf{c}_{k, l}}{\partial \hat{\theta}_{k, l}})}^2 + \frac{p^2}{4} {\hat{g}_{k, l}}^2 \vert {\mathbf{y}_{k}}^{(l)} \vert^{p-4} \odot \vert \mathbf{W}_k \frac{\partial \mathbf{c}_{k, l}}{\partial \hat{\theta}_{k, l}} \vert^2 \nonumber\\
           & \left. - \frac{p\cdot\hat{g}_{k, l}}{2} {{\mathbf{y}_{k}}^{(l)}}^{\text{H}} \mathbf{M} \mathbf{W}_k \frac{\partial^2 \mathbf{c}_{k, l}}{\partial \hat{\theta}_{k, l}^2}\right\}_,
    \end{align}
    \vspace{-0.3cm}
    \begin{align}
         \label{eq:derivative_3} \frac{\partial^2 S}{\partial \hat{\theta}_{k, l} \partial \hat{\varphi}_{k, l}} & = 2\Re \left\{\frac{p(p-2)}{4} {\hat{g}_{k, l}}^2 \left[ ({{\mathbf{y}_{k}}^{(l)}})^2 \odot \vert {{\mathbf{y}_{k}}^{(l)}} \vert^{p-4} \right] \odot  \right. \notag\\
         (\mathbf{W}_k \frac{\partial \mathbf{c}_{k, l}}{\partial \hat{\theta}_{k, l}}) & \odot (\mathbf{W}_k \frac{\partial \mathbf{c}_{k, l}}{\partial \hat{\varphi}_{k, l}}) 
          + \frac{p^2}{4} {\hat{g}_{k, l}}^2 \vert {\mathbf{y}_{k}}^{(l)} \vert^{p-4} \odot (\mathbf{W}_k \frac{\partial \mathbf{c}_{k, l}}{\partial \hat{\theta}_{k, l}}) \notag \\
          \odot (\mathbf{W}_k & \frac{\partial \mathbf{c}_{k, l}}{\partial \hat{\varphi}_{k, l}})^* + \left.\frac{p\cdot\hat{g}_{k, l}}{2} {{\mathbf{y}_{k}}^{(l)}}^{\text{H}} \mathbf{M} \mathbf{W}_k \frac{\partial^2 \mathbf{c}_{k, l}}{\partial \hat{\theta}_{k, l}\partial \hat{\varphi}_{k, l}}\right\}_.
    \end{align}
    
    After single refinement, we incorporate the newly detected and refined path into $\mathcal{P}_k$ and perform $R_c$ rounds of cyclic refinement. In cyclic refinement, a single refinement is performed upon each path while subtracting other detected paths. This multi-turbo-like refining procedure exploits the detection of the new path to facilitate more precise estimation of parameters of previous detected paths, thus overcoming the inaccuracy caused by the greedy nature of the OMP step. Additionally, to guarantee that the residual energy is non-increasing, a refinement is accepted only if the objective function $S$ decreases.  
    
    It is worth noting that the algorithm does not require the knowledge of the path number $L$. Therefore, we adopt a stopping criterion that the estimation process stops when the energy of the residual is smaller than the noise level known priorly as ${\Vert \mathbf{y}_k^{(l)} \Vert}_2^2 < \frac{{\Vert \mathbf{y}_k \Vert}_2^2}{{10}^{\frac{\text{SNR}}{10}} + 1}$, where SNR stands for signal-to-noise ratio, or the number of paths has reached the preset upper bound $L_\text{max}$. The steps of wNOMP are summarized in Algorithm \ref{code:wNOMP}.

    \vspace{-0.3cm}
    \begin{algorithm}[ht]
        \SetAlgoLined
        \caption{wNOMP via $\ell_p$-norm minimization}
        Initialization: $\mathbf{y}_k^{(0)} = \mathbf{y}_k$ for all $k = 1, 2, \hdots, K$\;
        \For{$k = 1 ~ \KwTo ~ K$}
        {
            $l=0$, $\mathcal{P}_k = \emptyset$\;
            \While{$l < L_{\text{max}}$}{
                \If{${\Vert \mathbf{y}_k^{(l)} \Vert}_2 < \frac{{\Vert \mathbf{y}_k \Vert}_2^2}{{10}^{\frac{SNR}{10}} + 1}$}
                {
                    return\;
                }
                
                coarsely detect new path $\mathbf{c}({\hat{\theta}}_{k, l}, {\hat{\varphi}}_{k, l}, {\tau}_{k, l}, \mathcal{N}_k)$ and $\hat{g}_{k, l}$ by  \eqref{eq:matching},  \eqref{eq:WLS} and  \eqref{eq:M}\;
                perform single refinement to $(\hat{\theta}_{k, l}, \hat{\varphi}_{k, l}, {\tau}_{k, l})$ by  \eqref{eq:Newton's_method}~-~\eqref{eq:derivative_3}\;
                
                $\mathcal{P}_k = \mathcal{P}_k \cup (\hat{\theta}_{k, l}, \hat{\varphi}_{k, l}, \hat{\tau}_{k, l})$\;
                \For{i = $1 ~ \KwTo ~ R_c$}
                {
                \For{$(\hat{\theta}_{k, p}, \hat{\varphi}_{k, p}, \hat{\tau}_{k, p}) \in \mathcal{P}_k$}
                {
                    $\mathcal{P}_k = \mathcal{P}_k \setminus (\hat{\theta}_{k, p}, \hat{\varphi}_{k, p},\hat {\tau}_{k, p})$\;
                    perform single refinement to $(\hat{\theta}_{k, p}, \hat{\varphi}_{k, p}, \hat{\tau}_{k, p})$ by  \eqref{eq:Newton's_method}~-~\eqref{eq:derivative_3}\;
                    $\mathcal{P}_k = \mathcal{P}_k \cup (\hat{\theta}_{k, p}^{\text{re}}, \hat{\varphi}_{k, p}^{\text{re}}, \hat{\tau}_{k, p}^{\text{re}})$\;
                }
                }
                use \eqref{eq:WLS} and  \eqref{eq:M} to update all path gain $\mathbf{\hat{g}}_k$ and the residual: $\mathbf{y}_k^{(l)} = \mathbf{y}_k - \mathbf{A} \mathbf{\hat{g}}_k$\;
                $l = l + 1$\;
            }
            recover $\mathbf{h}_k$ from $\mathcal{P}_k$ and $\mathbf{\hat{g}}_k$ by \eqref{eq:channel};
        }
    \label{code:wNOMP}
    \end{algorithm}
    \vspace{-0.3cm}

    \section{Convergence}
    The convergence of Algorithm \ref{code:wNOMP} is non-trivial to guarantee. The main difficulty is the fact that $\ell_p$-norm cannot induce an inner product when $p \neq 2$, which makes the results in \cite{NOMP} inapplicable. Here, we establish the convergence results using the \textit{$\ell_p$-correlation} proposed in \cite{l_p_correlation}.

    \begin{theorem}
        In each iteration of Algorithm \ref{code:wNOMP}, adding a new codeword $\mathbf{c}_{k, l}$ results in a decrease of the objective function $S$ in \eqref{eq:min_l_p_each_iteration} by the quantity of \textit{$\ell_p$-correlation} between $\mathbf{y}_k^{(l-1)}$ and $\mathbf{W}_k \mathbf{c}_{k, l}$, i.e.,
        \begin{align*}
            S(\mathbf{y}_k^{(l-1)}, \hat{g}_{k, l}, \mathbf{c}_{k, l}) \leq \Vert \mathbf{y}_k^{(l-1)} \Vert_p^p - c_p(\mathbf{W}_k \mathbf{c}_{k, l}, \mathbf{y}_k^{(l-1)}).
        \end{align*}
        The \textit{$\ell_p$-correlation} $c_p(\mathbf{W}_k \mathbf{c}_{k, l}, \mathbf{y}_k^{(l-1)})$ is strictly positive if $\hat{g}_{k, l} \neq 0$ (\textbf{Lemma 3} in \cite{l_p_correlation}). 
    \end{theorem}
    \begin{proof*}
        Since the detection step in Algorithm \ref{code:wNOMP} ensures $S(\mathbf{y}_k^{(l-1)}, \hat{g}_{k, l}, \mathbf{c}_{k, l}) \leq \Vert \mathbf{y}_k^{(l-1)} \Vert_p^p$, further applying \textbf{Theorem 1} in \cite{l_p_correlation} and the criterion that we only accept non-increasing refinement completes the proof. \hfill\qedsymbol
    \end{proof*}

    For the case when $\hat{g}_{k, l} = 0$, \cite{l_p_correlation} states that exhausting an overcomplete codebook will obtain a codeword such that $\hat{g}_{k, l} \neq 0$ and force the decrease of $S$. It requires a large codebook (reaching the DFT sampling limit) and is computationally impractical. However, the cyclic refinement and the path gain update in Algorithm \ref{code:wNOMP} can bypass this issue. Note that in the cyclic refinement step, the parameters of the previously detected paths will continuously be refined (despite that $\hat{g}_{k, l} = 0$), which can cause changes of the residue. Following this, the path gain update can re-allocate the coefficients to make $\hat{g}_{k, l} \neq 0$, which achieves a much more efficient "saddle point escaping" scheme.
    
    \section{Experiments}
        In this section, we present the experiment results of the proposed wNOMP algorithm under CGN and impulsive noise. We set the uplink carrier frequency $f_c = 30\text{GHz}$ and the bandwidth  $B = 1\text{GHz}$. The $N=128$ subcarriers are uniformly assigned to $K=8$ users (the algorithm also applies to other subcarrier assignation schemes). For the hybrid architecture at the BS, we set $M_v = M_h = 12,~M_\text{RF} = M_s = 32$. The channel is generated via \eqref{eq:channel} with $L=4, \theta_{k, l} \sim \mathcal{U}(0, \pi), \varphi_{k, l} \sim \mathcal{U}(-\pi, \pi), \tau_{k, l} \sim \mathcal{U}(0, 1/\Delta f), \bar{g}_{k, l} \sim \mathcal{CN}(0, 1)$. For the parameters of wNOMP and other grid-based methods, we set $N_{\hat{\theta}} = 4 M_v, N_{\hat{\varphi}} = 4 M_h, N_{\hat{\tau}} = 2 T, L_{max} = 10$.
        
        The normalized mean squared error (NMSE) is adopted to measure the performance of channel estimation
        \begin{align}
            \text{NMSE}_{\mathbf{h}} = \frac{1}{K} \sum_{k=1}^{K}\frac{\Vert \hat{\mathbf{h}}_{k} - \mathbf{h}_{k} \Vert_{\text{F}}^2}{\Vert \mathbf{h}_{k} \Vert_{\text{F}}^2}
        \end{align}
        
\begin{figure}[ht]
            \centering
            \includegraphics[scale=0.55]{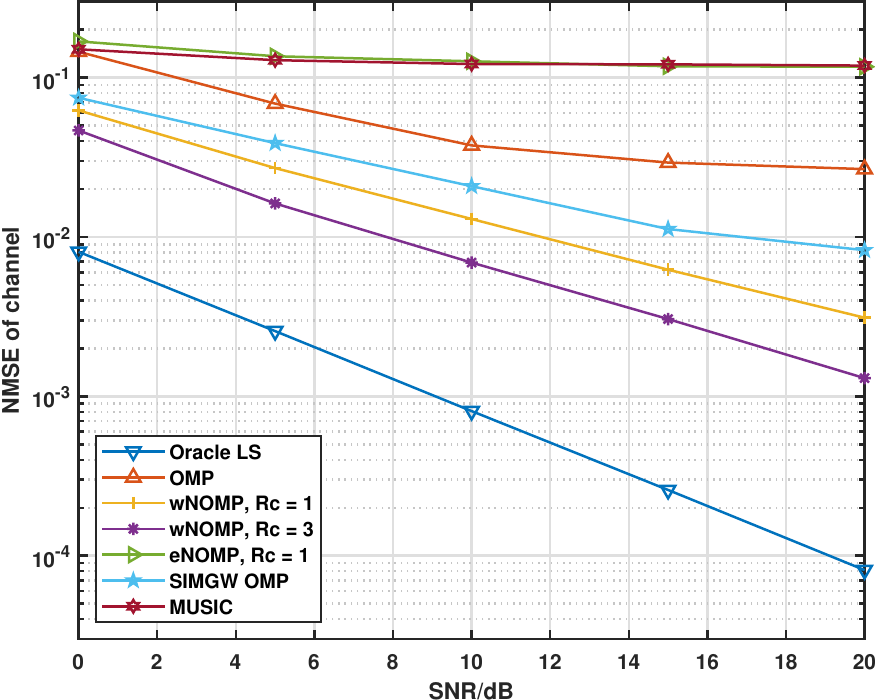}
            \caption{NMSE vs. SNR under CGN}
            \label{fig:all_comparison}
        \end{figure}

        In Fig. \ref{fig:all_comparison}, we compare the proposed wNOMP algorithm with the grid refinement method \textit{Simultaneous Iterative Multi-Gradient Weighted estimation exploiting common support} (SIMGW-OMP) proposed in \cite{SIMGW-OMP} and conventional OMP, \textit{MUltiple SIgnal Classification} (MUSIC) algorithm under CGN. These methods are all implemented into the wideband version for fair comparison. To investigate the phenonmon caused by beam squint, the eNOMP\cite{eNOMP} algorithm designed for the narrowband scenario is also added into benchmarks. The Oracle LS serves as a lower bound which assumes the parameter set $\mathcal{P}_k$ is already known. From Fig. \ref{fig:all_comparison}, it can be observed that the proposed wNOMP algorithm shows significant performance gain over other compressed-sensing-based methods. Particularly, both being gradient-based methods, wNOMP outperforms SIMGW-OMP as the Newtonized cyclic refinement step treats each component separately and achieves better accuracy.
        
\begin{figure}[ht]
            \centering
            \includegraphics[scale=0.55]{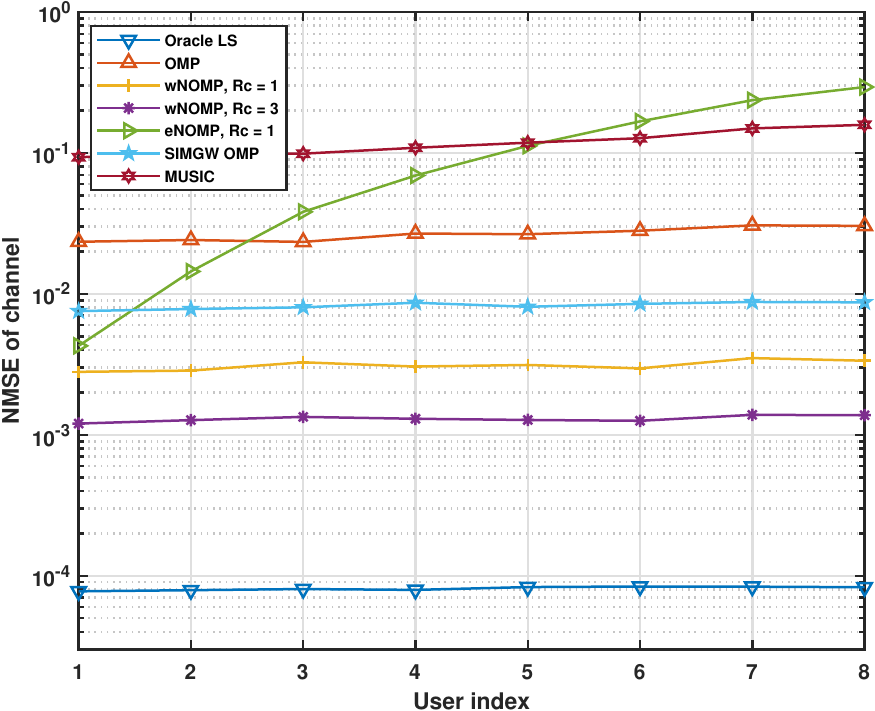}
            \caption{NMSE v.s. user index, SNR=20dB}
            \label{fig:NMSE_vs_user}
        \end{figure}

        In Fig. \ref{fig:NMSE_vs_user}, we focus on the accuracy of channel estimation of each user. Since the subcarriers of each user are aligned linearly across the total band, it can be expected that the performance of eNOMP degrades as the index of the user increases. This phenomenon leads to unfairness between the users and exhibits a typical effect of beam squint on conventional channel estimation algorithms. The proposed wNOMP algorithm maintains approximately the same performance by considering the beam squint effect.
        \vspace{-0.5cm}
        
\begin{figure}[ht]
            \centering
            \setlength{\abovecaptionskip}{0.cm}
            \includegraphics[scale=0.55]{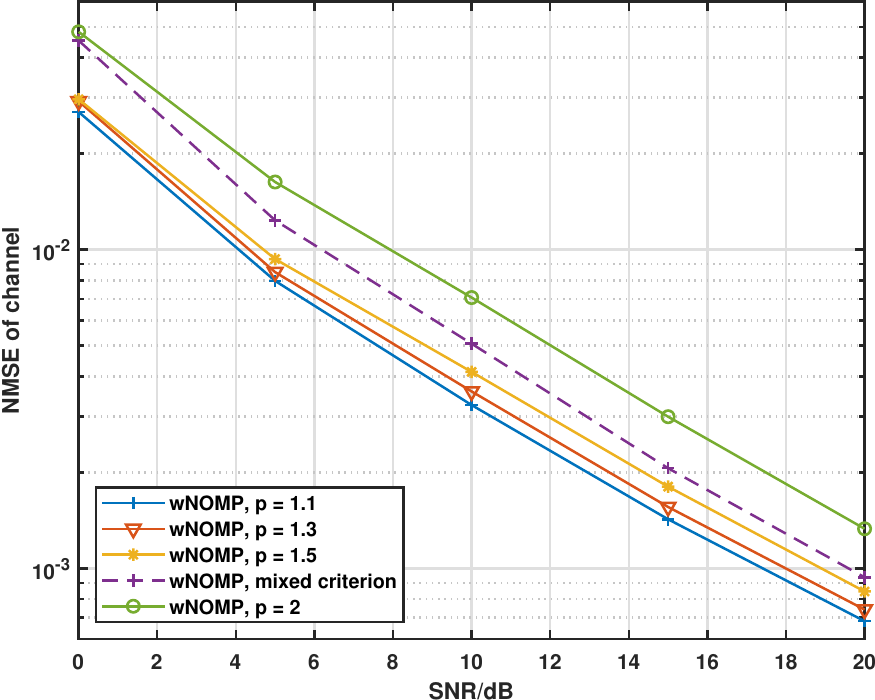}
            \caption{NMSE vs. SNR under Gaussian mixture noise}
            \label{fig:CGGN}
        \end{figure}

        Next, we study the performance of wNOMP under non-Gaussian noise, where the Gaussian mixture noise is adopted with two components as $(1-t)\cdot\mathcal{CN}(0, \sigma_1^2) + t\cdot\mathcal{CN}(0, \sigma_2^2)$. We set $t=0.1, \sigma_2 = 10\sigma_1$ that the second component can be viewed as a pulse generator. We use wNOMP with $1<p<2$ for heavy-tailed noise. Note that $p < 1$ is not considered because of the nonconvexity of the objective function when $p < 1$. Fig. \ref{fig:CGGN} shows that the minimum $\ell_p$-norm criterion is superior to the widely-used MMSE criterion with a range of $p$ from $1.1$ to $1.5$. To validate the effectiveness of grid refinement under minimum $\ell_p$-norm criterion, we compare the performances using minimum $\ell_p$-norm criterion ($p=1.1$) and using the mixed criterion (using $p=1.1$ IWLS and MMSE Newton's method). It turns out that the former outperforms the latter which proves the effectiveness of Newton's method under the minimum $\ell_p$-norm criterion. As for the selection of $p$, since the Gaussian mixture noise used here contains an impulsive tail and can be better captured by CGGN with a heavier tail, the performance of wNOMP is better when $p$ takes a smaller value. In experiments, we have also observed that smaller $p$ averagely takes more iterations to converge due to the optimization property of $\ell_p$-norm.  

        \vspace{-0.3cm}
        \begin{table}[ht]
            \centering
            \caption{Complexity of Algorithms}
            \begin{tabular}{cc}
               \toprule
               Algorithm & Computational Complexity\\
               \midrule
               OMP / SIMGW-OMP & $\mathcal{O}(M_sM_vM_h T ( N_{\hat{\theta}} N_{\phi} N_{\tau} + T L^2))$\\
               wNOMP / eNOMP & $\mathcal{O}(M_sM_vM_h T ( N_{\hat{\theta}} N_{\phi} N_{\tau} + T L^2 R_c))$\\
               MUSIC & $\mathcal{O}(M_sM_vM_h T ( N_{\hat{\theta}} N_{\phi} N_{\tau} + T L^2) + M_s^3T^3)$\\
               \bottomrule
               \label{table:complexity}
            \end{tabular}
        \end{table}
        \vspace{-0.3cm}
                
        The complexity results of these methods are presented in Table \ref{table:complexity}. From Table \ref{table:complexity}, it can be noticed that the cyclic refinement introduces an extra complexity by $\mathcal{O}(M_sM_vM_hT^2LR_c)$. If we restrict $R_c$ to a small value, e.g., $R_c \leq 5$ (we observe that $R_c = 3$ already works well enough experimentally), the extra computational burden can be ignored since the number of paths $L$ is usually small in mmWave communication systems.
       
    \section{Conclusion}
       In this paper, a robust wideband channel estimation algorithm based on coarse-to-fine detection has been proposed to tackle the beam squint effect. By using the minimum $\ell_p$-norm criterion, the robustness against the impulsive noise can be improved. When $p = 2$, it reduces to the conventional MMSE criterion, which is optimal for Gaussian noise. Convergence results are also established for the proposed method. Experiments have been conducted to validate the advantage of the proposed method over other representative methods.


\appendix
    In Appendix, we give a detailed derivation of the fixed point expression of path gain vector $\mathbf{\hat{g}}_k$, along with derivatives of the objective function $S$ over parameters $\hat{\theta}_{k, l}, \hat{\phi}_{k, l}$, and $\hat{\tau}_{k, l}$. Specifically, we give the derivation of $\frac{\partial S}{\partial \hat{\theta}_{k, l}}, \frac{\partial^2 S}{\partial {\hat{\theta}_{k, l}}^2}$ and $\frac{\partial^2 S}{\partial \hat{\theta}_{k, l} \partial \hat{\phi}_{k, l}}$. The remaining first-order and second-order derivatives can be directly extended.
    
    Our goal is to minimize the objective function
    \vspace{-0.3cm}
    \begin{align}
        J = & \Vert \mathbf{y}_k - \mathbf{W}_k\sum_{l=1}^{L_c} \hat{g}_{k, l} \mathbf{c}(\hat{\theta}_{k, l}, \hat{\varphi}_{k, l}, \hat{\tau}_{k, l}, \mathcal{N}_k) \Vert_p^p \nonumber\\
          = & \Vert \mathbf{y}_k - \mathbf{W}_k\mathbf{X} \mathbf{\hat{g}}_k \Vert_p^p. \label{eq:J}
    \end{align}
    The optimal $\mathbf{\hat{g}}_k$ can be obtained by letting $\frac{\partial J}{\partial \mathbf{\hat{g}}_k} = \mathbf{0}$. Define $\epsilon_i =  \vert {(\mathbf{y}_k)}_i - \sum_{l=1}^{L_c} (\mathbf{W}_k\mathbf{X})_{i, l} \hat{g}_{k, l} \vert$, it is equivalent to
    \begin{align}
         &\sum_{i=1}^{M_vM_hT} {\epsilon_i}^{p-2} (\mathbf{W}_k\mathbf{X})_{i, l} {(\mathbf{y}_k)_i}^*\nonumber\\
         & = \sum_{i=1}^{M_vM_hT} {\epsilon_i}^{p-2} (\mathbf{W}_k\mathbf{X})_{i, l} \sum_{l=1}^{L_c} (\hat{g}_{k, l} (\mathbf{W}_k\mathbf{X})_{i, l})^*.
    \end{align}
    Donate $\mathbf{M} = \text{diag}\left\{ {\vert \mathbf{y}_k - \mathbf{W}_k\mathbf{X} \mathbf{\hat{g}}_k \vert}^{p-2} \right\}$ as the diagonal weighting matrix and the above equation turns into
    \begin{align}
        \mathbf{\hat{g}}_{k} = {((\mathbf{W}_k\mathbf{X})^{\text{H}}\mathbf{M}\mathbf{W}_k\mathbf{X})}^{-1} (\mathbf{W}_k\mathbf{X})^{\textbf{H}}\mathbf{M}\mathbf{y}_k .
        \label{eq:fixed_point_expression}
    \end{align}
    \eqref{eq:fixed_point_expression} gives the fixed-point-type requirement that the optimal $\mathbf{\hat{g}}_k$ satisfies. It can be readily translated into the IWLS algorithm in the paper.

    \vspace{2\baselineskip}
    To obtain derivatives over $\hat{\theta}_{k, l}, \hat{\phi}_{k, l}$, and $\hat{\tau}_{k, l}$, first we obtain the derivatives over the compressed codeword $\mathbf{x} = \mathbf{W}_k \mathbf{c}_{k, l}$, where $\mathbf{c} \in \mathcal{C}(\hat{\mathbf{\theta}}, \hat{\mathbf{\varphi}}, \hat{\mathbf{\tau}}, \mathcal{N}_k)$: 
        \begin{align}
            &\label{eq:derivative_x}\frac{\partial S}{\partial \mathbf{x}} = \frac{-p \cdot \beta \cdot \mathbf{M} \cdot {\mathbf{y}_k^{(l)}}^*}{2} ,\\
            &\label{eq:derivative_x_x}\frac{\partial^2S}{\partial\mathbf{x}^2} = \frac{p(p-2)}{4} \cdot \beta^2 \cdot {({{\mathbf{y}_k^{(l)}}^*})}^2 \otimes {\vert \mathbf{y}_k^{(l)} \vert}^{s-4} ,\\
            &\label{eq:derivative_x_x_conj}\frac{\partial^2S}{\partial\mathbf{x}\partial \mathbf{x}^{*}} = \frac{p^2}{4} \cdot \beta^2 {\vert \mathbf{y}_k^{(l)} \vert}^{s-2}.
        \end{align}
    Applying the chain rule, we have
    \begin{align}
        \label{eq:derivative_theta}\frac{\partial S}{\partial \hat{\theta}_{k, l}} & = 2\Re\left\{{(\frac{\partial S}{\partial \mathbf{x}})}^{\text{T}} \frac{\partial \mathbf{x}}{\partial \hat{\theta}_{k, l}}\right\},
    \end{align}
    \vspace{-0.3cm}
    \begin{align}
        \label{eq:derivative_theta_theta}\frac{\partial^2 S}{\partial {\hat{\theta}_{k, l}}^2} & = 2\Re\left\{{(\frac{\partial^2S}{\partial\mathbf{x}^2})}^{\text{T}} {(\frac{\partial \mathbf{x}}{\partial \hat{\theta}_{k, l}})}^2 \right. \notag
        \\& + {(\frac{\partial^2S}{\partial\mathbf{x}\partial \mathbf{x}^*})}^{\text{T}} {\vert\frac{\partial \mathbf{x}}{\partial \hat{\theta}_{k, l}}\vert}^2 + {(\frac{\partial S}{\partial\mathbf{x}})}^{\text{T}} (\frac{\partial^2 \mathbf{x}}{\partial \hat{\theta}_{k, l}^2}),
    \end{align}
    \vspace{-0.3cm}
    \begin{align}
        \label{eq:derivative_theta_phi} & \frac{\partial^2 S}{\partial \hat{\theta}_{k, l} \partial \hat{\phi}_{k, l}} = 2\Re\left\{{(\frac{\partial^2S}{\partial\mathbf{x}^2})} \odot {(\frac{\partial \mathbf{x}}{\partial \hat{\theta}_{k, l}})} \odot {(\frac{\partial \mathbf{x}}{\partial \hat{\phi}_{k, l}})} \right.\notag \\ & \left.+{(\frac{\partial^2S}{\partial\mathbf{x}\partial\mathbf{x}^*})} \odot {\frac{\partial \mathbf{x}}{\partial \hat{\theta}_{k, l}}} \odot {(\frac{\partial \mathbf{x}}{\partial \hat{\phi}_{k, l}})}^{*}
        {(\frac{\partial S}{\partial\mathbf{x}})}^{\text{T}} (\frac{\partial^2 \mathbf{x}}{\partial \hat{\theta}_{k, l} \partial \hat{\phi}_{k, l}}) \right\}.
    \end{align}
    Plug  \eqref{eq:derivative_x},  \eqref{eq:derivative_x_x},  \eqref{eq:derivative_x_x_conj} into  \eqref{eq:derivative_theta},  \eqref{eq:derivative_theta_theta},  \eqref{eq:derivative_theta_phi} and we obtain the complete derivatives over $\theta, \phi, \tau$:
   \begin{align}
        \centering
        \frac{\partial S}{\partial \hat{\theta}_{k, l}} & = 2\Re{\left\{ \frac{p\cdot\hat{g}_{k, l}}{2} {{\mathbf{y}_{k}}^{(l)}}^{\text{H}} \mathbf{M} \mathbf{W}_k \frac{\partial \mathbf{c}_{k, l}}{\partial \hat{\theta}_{k, l}}\right\}}_, 
    \end{align}
    \vspace{-0.3cm}
    \begin{align}
        \frac{\partial^2 S}{\partial \hat{\theta}_{k, l}^2} & = 
        2\Re \left\{\frac{p(p-2)}{4} {\hat{g}_{k, l}}^2 \left[ ({{\mathbf{y}_{k}}^{(l)}}^{\text{H}})^2 \odot \vert {{\mathbf{y}_{k}}^{(l)}}^{\text{H}} \vert^{p-4} \right] \right. \notag\\
           & {(\mathbf{W}_k \frac{\partial \mathbf{c}_{k, l}}{\partial \hat{\theta}_{k, l}})}^2 + \frac{p^2}{4} {\hat{g}_{k, l}}^2 \vert {\mathbf{y}_{k}}^{(l)} \vert^{p-4} \odot \vert \mathbf{W}_k \frac{\partial \mathbf{c}_{k, l}}{\partial \hat{\theta}_{k, l}} \vert^2 \nonumber\\
           & \left. - \frac{p\cdot\hat{g}_{k, l}}{2} {{\mathbf{y}_{k}}^{(l)}}^{\text{H}} \mathbf{M} \mathbf{W}_k \frac{\partial^2 \mathbf{c}_{k, l}}{\partial \hat{\theta}_{k, l}^2}\right\}_,
    \end{align}
    \begin{align}
         \frac{\partial^2 S}{\partial \hat{\theta}_{k, l} \partial \hat{\varphi}_{k, l}} & = 2\Re \left\{\frac{p(p-2)}{4} {\hat{g}_{k, l}}^2 \left[ ({{\mathbf{y}_{k}}^{(l)}})^2 \odot \vert {{\mathbf{y}_{k}}^{(l)}} \vert^{p-4} \right] \odot  \right. \notag\\
         (\mathbf{W}_k \frac{\partial \mathbf{c}_{k, l}}{\partial \hat{\theta}_{k, l}}) & \odot (\mathbf{W}_k \frac{\partial \mathbf{c}_{k, l}}{\partial \hat{\varphi}_{k, l}}) 
          + \frac{p^2}{4} {\hat{g}_{k, l}}^2 \vert {\mathbf{y}_{k}}^{(l)} \vert^{p-4} \odot (\mathbf{W}_k \frac{\partial \mathbf{c}_{k, l}}{\partial \hat{\theta}_{k, l}}) \notag\\
          \odot (\mathbf{W}_k & \frac{\partial \mathbf{c}_{k, l}}{\partial \hat{\varphi}_{k, l}})^* + \left.\frac{p\cdot\hat{g}_{k, l}}{2} {{\mathbf{y}_{k}}^{(l)}}^{\text{H}} \mathbf{M} \mathbf{W}_k \frac{\partial^2 \mathbf{c}_{k, l}}{\partial \hat{\theta}_{k, l}\partial \hat{\varphi}_{k, l}}\right\}_.
    \end{align}

\end{document}